\documentclass{iaus}
\usepackage{graphicx}

\title[Ground-Based Searches for Transiting Planets] 
{Ground-Based Photometric Searches for Transiting Planets}

\author[T. Mazeh]   
{Tsevi Mazeh} 
\affiliation{Wise Observatory, Tel Aviv University, Tel Aviv 69978,
    Israel. \\ email: {\tt mazeh@post.tau.ac.il 
\\ Present temporary address: Radcliffe Institute for
Advanced Studies at Harvard} \\[\affilskip]
}

\pubyear{2008}
\volume{253}  
\pagerange{1--10}
\jname{Transiting Planets}
\editors{TBD}
\begin{document}

\maketitle

\begin{abstract}

This paper reviews the basic technical characteristics of the
ground-based photometric searches for transiting planets, and
discusses a possible observational selection effect. I suggest that
additional photometric observations of the already observed fields
might discover new transiting planets with periods around 4--6
days. The set of known transiting planets support the intriguing
correlation between the planetary mass and the orbital period
suggested already in 2005.


\end{abstract}


\firstsection
\section{Introduction}

The first known transiting planet, HD\,209458b (\cite[Charbonneau et
al. 2000]{char00}; \cite[Henry et al. 2000]{hen00}), was discovered as
a extra-solar planet by the radial-velocity (RV) technique
(\cite[Mazeh et al. 2000]{maz00}; \cite[Henry et
al. 2000]{hen00}). This planet attracted much interest because the
observations of the transit in different wavelengths opened up a wide
window through which we could study for the first time some planetary
features of an extra-solar planet. This includes measuring the
planetary mass and radius, studying the planetary envelope
(Charbonneau et al. 2002; Vidal-Majar et al. 2003; 2003) and even
deriving the planetary direction of motion relative to the rotation of
its parent star (Queloz et al. 2000).

Since then two additional interesting transiting planets were found by
ground-based observations {\it after} their discovery as planets by
the RV technique. HD\,149026 (\cite[Sato el al. 2005]{sato05}) with
its relatively small radius is one of the very few planets that
probably have a metallic, rocky core, and Gls\,436 (\cite[Butler et
al.]{}; \cite[Maness et al]{}; \cite[Gillon et al.]{}) is the only
transiting planet known with a Neptunian mass and radius.

Although these three systems contributed substantially to our
knowledge of extra-solar planets, the bulk of the information we
acquired about planetary radii and masses comes from the transiting
planets that were discovered by systematic ground-based photometric
searches. For these planets, the order of detection is reversed --- we
first detect transiting planet candidates by photometry and only then
confirm their planetary nature by RV {\it follow-up} observations.

The photometric search for transiting planets has two types of
drawbacks. The first type is associated with the fact that the
detectability of transiting planets is limited to only those planets
whose orbits cross the disc of their parent stars, as seen from our
line of sight. For a given stellar radius, $R_*$, planetary radius,
$r_p$, and semi-major axis $a$, the fraction of planets with circular
orbits and random orientations that transit their parent stars is
\begin{equation}
Prob(transit)=\frac{R_*+r_p}{a} \ .
\end{equation}
\label{geometry}
Therefore, the fraction of transiting planets is about 10\% at most,
even for the hot-Jupiters with periods of the order of 3 days. The
fraction goes down to 1\% for planets around solar-type stars with
periods of about 100 days, limiting the search for transiting
planets to short-period planets only. 

\begin{figure}[b]
\begin{center}
\includegraphics[width=4in, height=3in]{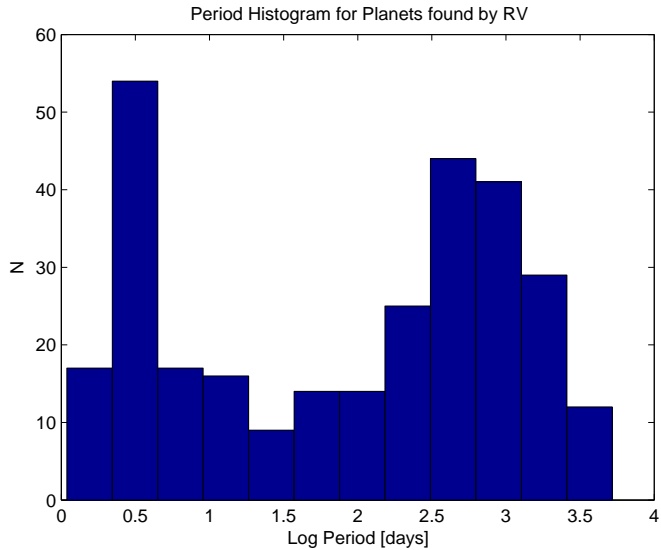}
 \caption{The period distribution of the planets found by
   RV observations.}
   \label{fig_per_his}
\end{center}
\end{figure}

Fortunately, the short-period planets pose quite a few intriguing open
questions, including migration and stopping mechanism, tidal
interaction and heating by the insolation of the close by
star. Moreover, the short-period planets compose a substantial part of
the known population of the extra-solar planets, as can be seen from
Figure~1, which presents an histogram of the known planets found by
the RV technique. One can see that the observed frequency rises when
going from 10 to 100 days, indicating that the longer the period the
more planets we are about to find by RV observations (the drop of
frequency for periods longer than 100 days is due to RV selection
effects.) In addition, the histogram shows one pronounced peak
consisting of planets with periods at about 3 days. This peak is
probably caused by the migration mechanism, which apparently prefers
to shrink the planet orbits into periods of about 3 days. These
planets are the main candidates for being transiting planets. The
photometric search, therefore, enables us to study in details this
intriguing subset of extra-solar planets.

One other type of drawback of the photometric search for transiting
planets is associated with the discovery technique. First, because of
the relatively low frequency of the short-period planets and the low
percentage of planets that transit their parent stars, one needs to
observe many hundreds of solar-type stars in order to discover a
transiting planet. In addition, in any field on the sky the solar-type
stars are diluted by many F main-sequence and by giant stars, which
cannot possibly show detectable transits. So, on the average, one has
to follow of the order of 10,000 stars in order to find one transiting
planet.

Second, the expected transit depth is of the order of only one percent
and its phase duration is short, of the order of a few hours for a
period of 3 days. Therefore, we need many accurate measurements in
order to discovery the transit.  Moreover, variability induced by
stellar activity together with observational errors can easily mask
the transit minute modulation. Third, many eclipsing binaries, some of
which are blended with one more star in the obtained image of the
field, disguise as transiting planets. Only RV observations can lift
their mask and discover their true nature. Therefore, the photometric
searches depend heavily of RV follow-up observations.

Despite all the drawbacks of the photometric approach, it proved to be
very efficient in finding transiting planets, yielding about 30 of
them as of May 2008. This is so because newly available CCD technology
enables us to perform efficient photometric surveys. The size of the
new CCDs, some with 4K$\times$4K pixels, enables us to follow the
brightness of tens or even hundreds of thousands stars per exposure,
the efficiency of the CCDs allows photometric measurements with a
cadence of the order of a few minutes, and the precision of the new
devices could reach a few millimag for a measurement of a stellar
brightness.

However, it seems that the advanced technology of the new CCDs is not
enough. The very first years of the photometric systematic searches
that followed the discovery of HD\,209458b in 2000 yielded much less
transiting planets than expected (e.g., Horne, 2003), except the
outstanding success of the OGLE team (e.g., Udalski et al. 2002a,b;
2003; 2004; 2005; Pont et al. 2007a; 2008). Apparently, the proper
analysis of the data might also have a role in the success of the
photometric search (e.g., Pont et al. 2006; 2007b). After some
algorithms to search for the transits (e.g., BLS ---- Kov\'acs 2002)
and to clean the data (e.g., SysRem (Tamuz et al. 2005) and TFA (Bakos
et al. 2005)) were developed, and when new specifically build small
telescopes became available for routine observations, the yield of the
photometric searches became substantial.

This paper reviews the technical details of present and
planned systematic photometric searches (Section~\ref{review}) and
points to a possible selection effect hindering the detection of
transiting planets with periods in the range of 4--6 days
(Section~\ref{selection_effects}). Finally, Section~\ref{mass_period}
reviews the accumulated evidence for the intriguing correlation between
the mass and the period of the short-period planets.

%
%
\section{Present and future photometric systematic searches}
\label{review}
%
%
This section reviews some basic features of present and planned
systematic photometric searches for transiting planets, in order to
give the grand picture of this endeavor.  Making every effort to bring
the correct figures relevant to each project, the review is based on
correspondce I have had with members of each team.

A short summary is given in Table~1, which lists the name of the
project, diameter of the telescope(s) used, size of individual field
of view (FoV) observed (in square-degrees), number of CCD pixels used
per field, number of telescopes operated by the project, number of
fields observed so far, averaged number of measurements acquired so
far per field and an estimate of the total number of stars observed so
far to a precision of 1\% per measurement. Finally, the last column brings
the number of planets discovered so far by the project. The number of
published transiting planets published is given first and then, in
parenthesis, the number of planets announced, as of May 2008.
 
\begin{table}
  \begin{center}
  \caption{The present and future systematic photometric searched for
  transiting planets.}
  \label{tab1}
 {
  \begin{tabular}{|c|c|c|c|c|c|c|c|c|c|}\hline

& Tel. & FoV& Pixel& Tel.& FoV& Meas.& Stars& mag& Planets \\
&      & (Sq. &    &     &       &    & Observed&range& discovered\\
& (cm) & deg) &(\#) & (\#) & (\#) & (\#) & (\#) & (V) & (\#)   \\
\hline
WASP N & 11.1& 61 & 4M & 8 & 200 & 6,000 &1,000,000 & 13 & 3+5\\
WASP S & 11.1& 61 & 4M & 8 & 120 &10,000 &  500,000 & 13 &  2 \\
XO     &10.\ & 51 & 2M & 2 &  90 & 3,000 &  250,000 & 12 & 3+2\\ 
HATnet & 11  & 100& 16M& 6.5& 50 & 5,000 &  500,000 & 12 & 7 \\
TrES   & 10  & 36 & 4M & 3  & 19 & 8,000 &   50,000 & 14 & 4 \\
\hline
OGLE   & 130 & 0.36& 64M& 1 & 20 & 1000  & 500,000  & 16 & 7\\
ANU Lupus&100& 0.66& 60M& 1 &  1 & 3000  &  15,000  & 17 & 1\\
\hline
HAT S  & 11 & 100 & 16M& 24  \\
BEST   & 19.5& 9.6& 4M & 1  \\
BEST II& 25  & 2.9&16M & 1  \\
\hline
LAIWO            & 100 & 1  & 64M & 1 \\
ANU skymapper & 130 &5.7 & 270M & 1 \\
Pan-STARRS       & 180 & 7  &1400M & 1  \\
\hline
  \end{tabular}
  }
 \end{center}
\end{table}


The upper part of the table reviews the presently active projects. It
includes five small-telescope projects, which use telescopes with
diameters of the order of 10 cm: WASP N \&S (e.g., Pollacco et
al. 2006) XO (McCullough et al. 2005), HATnet (Bakos et al. 2002;
2004), and TrES (e.g., Alonso et al. 2004). The table lists WASP N (north)
and WASP S (south) in two separate lines, although they are two parts
of the same project. Note that the two parts of the WASP project are
covering together about 20,000 square-degrees already, which is about half of
the whole sky.

The upper part of the table also lists OGLE (Udalski et al. 2002) and
ANU Lupus (Weldrake et al. 2007), two projects that use 1-m class
telescopes for some of their observing nights, and therefore their
field of view is of the order of half square-degree. This should be
compared with the large FoV of the small telescopes, which are of the
order of hundred times larger. Note also that the 1-m class telescopes
detection range is V=16--17 mag, 40--100 fainter than the transits
found by the small telescopes.

The second part of the table lists the planned projects, which include
one project with small telescopes --- HAT S, which will have extremely
large observing power; two projects with 
intermediate telescopes, BEST and BEST II (Kabath et al. 2007), with
telescopes slightly larger than the small ones; two projects with 1-m
class telescopes --- LAIWO (Afonso et al. 2006) and ANU skymappers
(Bayliss \& Sackett 2007); and finally the ambitious project ---
Pan-STARRS (e.g., Afonso \& Henning 2007) with a 2-m class telescope.

The small telescopes were built and are operated for detecting
transiting planets only, while the larger telescopes were and are used
for other projects too. As a consequence, the small telescope
observing time is devoted almost completely to the search for
transiting planets. Together with their large FoV, this will lead to
the coverage of the whole sky in the very near future. We therefore
anticipate that the small-telescope projects probably will go back to
the already observed fields and add many more observations. As will be
suggested in the next section, additional observations might lead to
the discovery of more transiting planets.

\section{A possible selection effect of the ground-based
  small-telescope searches}
\label{selection_effects}

\begin{figure}[b]
\begin{center}
\includegraphics[width=4in]{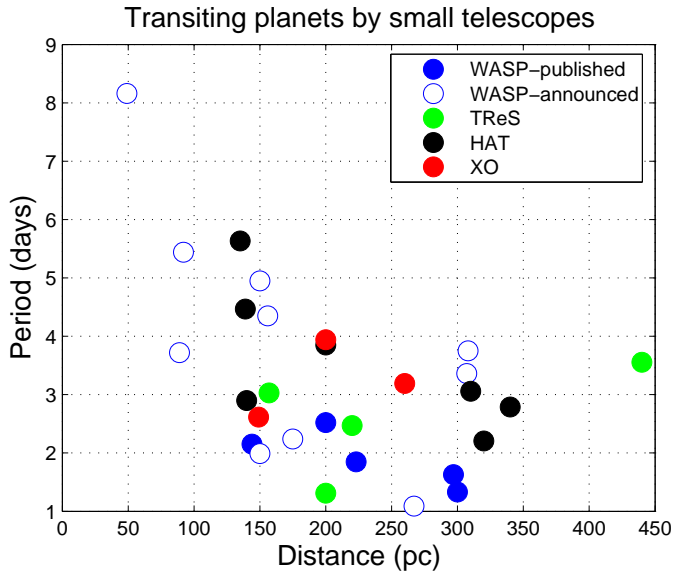}
 \caption{The orbital period of the transiting planets discovered by
small telescopes as a function of their distance from the Sun. [See
electronic version of the paper for the coloured figure.]}
   \label{P_dist}
\end{center}
\end{figure}

This section discusses the set of transiting planets discovered by the
ground-based small-telescope projects and considers a possible
selection effect acting against discovering planets with long periods
(see discussion by Gaudi et al. 2005). A suggestion that such a
selection effect is in action might be found in Figure~\ref{P_dist},
which shows the orbital period of the transiting planets discovered by
the small telescopes as a function of their estimated distance from
the Sun. Although the distances of the individual stars are not very
well known, we assume that the general trend that appears in the
figure is correct. Assuming the distribution of planets in the Solar
neighbourhood is constant, the dependence seen in Figure~\ref{P_dist}
can be attributed only to observational selection effects.

The obvious assumption would be that the selection effect observed in
Figure~\ref{P_dist} is due to the fact that the photon noise, which is
one of the major factors of the observational noise for small
telescopes, is larger for more distant, and therefore fainter,
systems. For the more distant transiting planets we need more
observations to occur within the transit in order to detect the small
periodic drop of the stellar brightness. On the other hand, it is more
difficult to discover transits with longer periods because of two
effects. First, the number of individual transits actually observed
for a given observational time span is inversely proportional to the
length of the period. Second, the phase of the transit, $f$, which can
be written for a circular orbit as
\begin{equation}
f=\frac{R_*}{\pi a} \ ,
\end{equation}
is shorter for longer periods.  Therefore, for a given number of
observational points, the number of points within the transit is a
decreasing function of the period. Both effects turn transiting
planets with longer periods to be more difficult to discover, and
therefore their detection threshold to be brighter.

\begin{figure}[b]
\begin{center}
\includegraphics[width=4in]{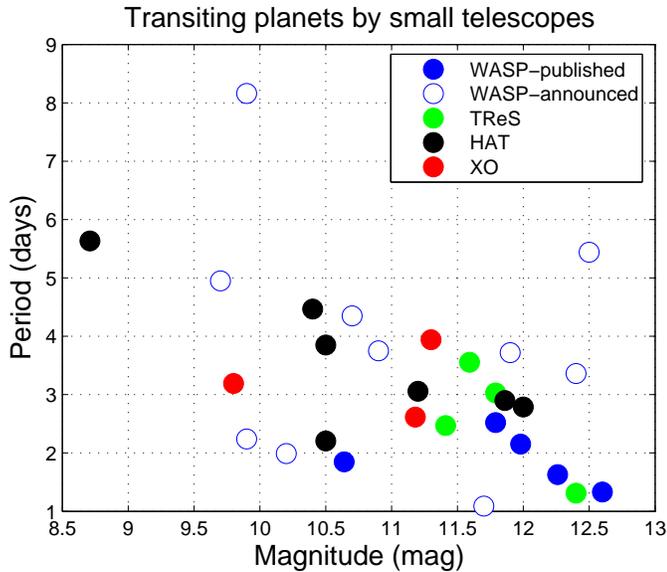}
 \caption{The orbital period of the transiting planets discovered by 
small telescopes as a function of the their parent star brightness. [See
electronic version of the paper for the coloured figure.]}
   \label{P_mag}
\end{center}
\end{figure}

This can also be seen in Figure~\ref{P_mag}, which shows
the orbital periods of the detected transiting planets as a function of
the magnitude of their parent stars. The very clear dependence supports
the assumption about the selection effect.

One possible prediction that might be drawn from the suggested
selection effect is that the small-telescope projects might find
additional transiting planets if they go back to the fields already
observed and increase substantially the number of observation per
star. The additional observations might fill up, for example, the
parameter space in Figure~\ref{P_mag} with period range between 2 and
5 days for stars fainter than 12 mag.

%
%
\section{The mass-period relation for the very short-period planets}
\label{mass_period}
%
%

\begin{figure}[b]
\begin{center}
\includegraphics[width=4.5in, height=3in]{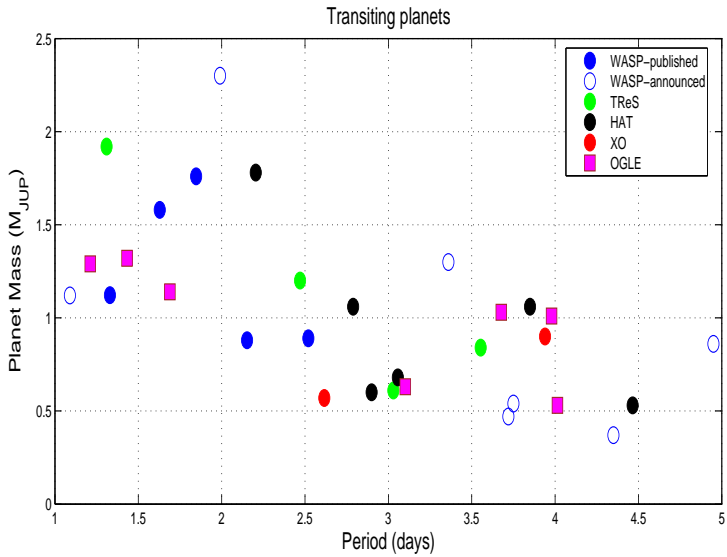}
 \caption{The mass of the transiting planets as a function of their
orbital period. [See electronic version of the paper for the coloured
figure.] }
   \label{mass_per}
\end{center}
\end{figure}

Mazeh et al. (2005) already noticed a correlation between the
mass of the transiting planets and their period (see also Gaudi et
al. 2005). As that suggestion was based on only six transiting
planets known at that time, it is of some interest to revisit the
mass-period diagram and see if the correlation still holds. To do that
we plotted in Figure~\ref{mass_per} the derived masses of the known
transiting planets as a function of their orbital periods. The figure
shows all transiting planets discovered by photometry, including the
ones detected by OGLE. One can see clearly that the correlation still
holds.

The astrophysics behind this correlation is not clear. One
could argue that if a planet gets too close to its parent star, the
stellar insolation tends to evaporate the planetary atmosphere.
Therefore, only planets with large enough masses and therefore large
surface gravities can survive at short distances from their parent
stars. However, this argument only explains why there are no planets
with small masses {\it and} short periods, but cannot explain why
there are almost no planets with large masses {\it and} long periods.

\begin{figure}[b]
\begin{center}
\includegraphics[width=4in]{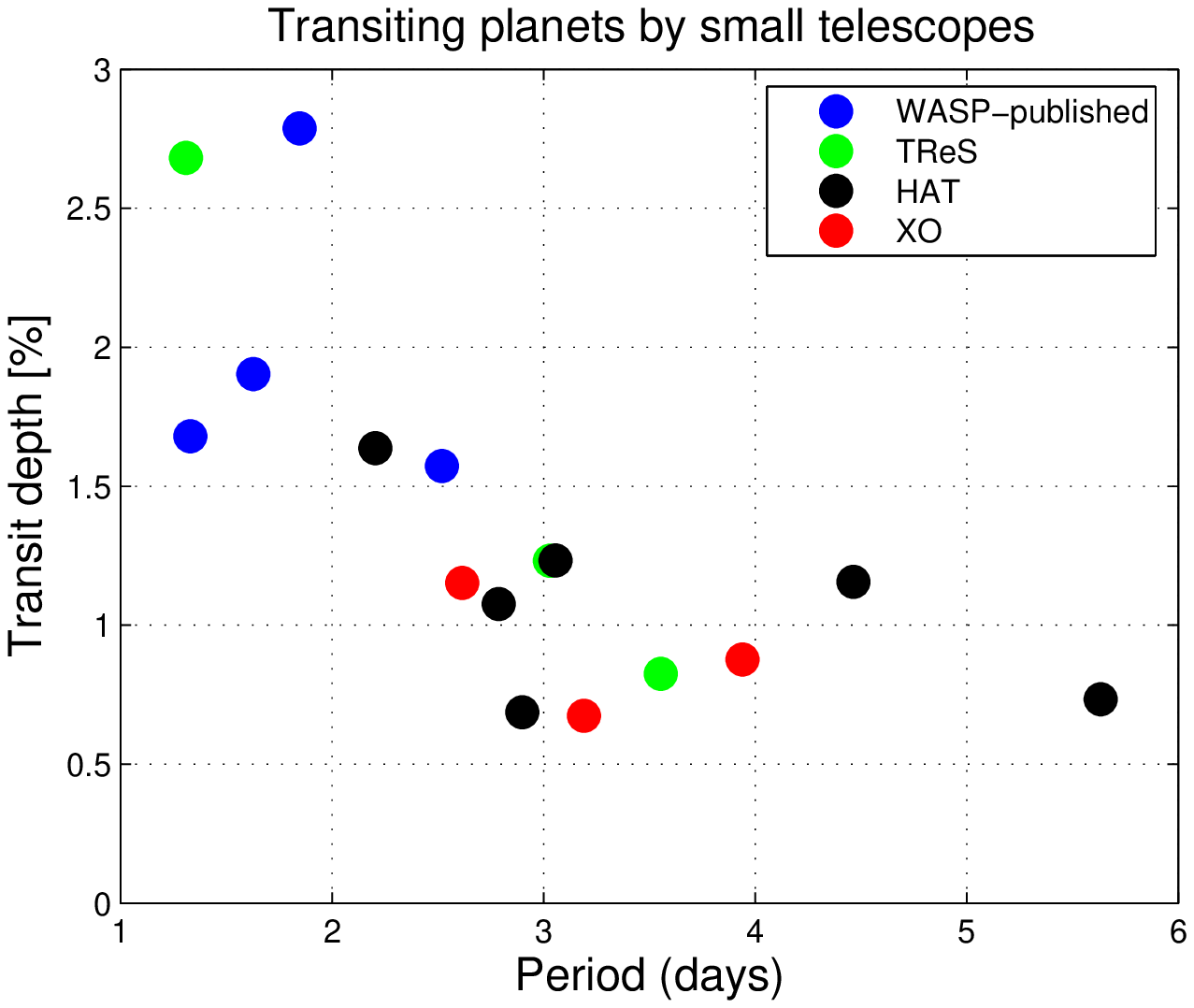}
 \caption{The depth of the transits of the planets discovered by 
small telescopes as a function of their orbital period. [See
electronic version of the paper for the coloured figure.]}
   \label{depth_per}
\end{center}
\end{figure}

One could see the same effect by looking at the dependence of the
depth of the transit on the orbital period, as done in
Figure~\ref{depth_per}. The figure suggests that the transit depths
are shallower for long periods.  This is probably a real effect,
because detection considerations alone predict the opposite dependence,
as discussed in the previous section. Therefore, the fact that the
known long-period transiting planets have shallower depth can be
explained only if the ratio between the stellar radius and the
planetary radius depends on the planetary period.  This has probably
to do with the mass-period correlation, which induces a radius-period
dependency.

\section{Discussion}

\begin{figure}[b]
\begin{center}
\includegraphics[width=4in]{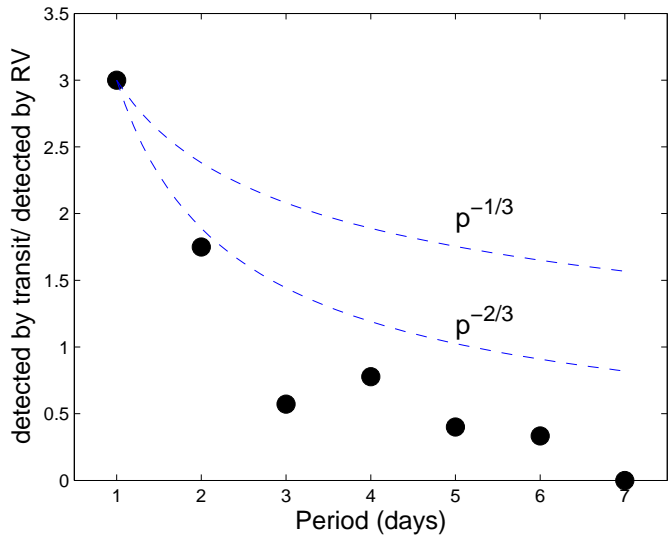}
 \caption{The ratio of transiting planets discovered by photometry
 to the ones discovered by the RV technique.}
   \label{ratio}
\end{center}
\end{figure}

The ground-based systematic photometric searches for transiting
planets have proven themselves to be observational projects with high
yield. With relatively small investment of resources they are
producing a flux of transiting extra-solar planets. To estimate the
efficiency of the photometric searches, it might be of interest to
compare the number of planets discovered by photometry with the number
of planets discovered by RV technique (Gaudi et al. 2005). This is
done in Figure~\ref{ratio}, where the number of planets discovered by
photometry is {\it divided} by the number of planets detected by RV
observations is plotted in period bins of one day. In the first bin,
which presents the ratio of planets found in the range of [0.5, 1.5]
days, this ratio is 3, which indicates that the photometric search is
more effective than the RV technique. However, this ratio drops
dramatically for longer periods. As can be seen in the figure, this
ratio is about 0.5 for periods longer than 2 days.

We do expect this ratio to fall off for longer periods because of the
geometrical effect (see Equation (1.1)). However, it seems as if the
drop is sharper than expected. To see that this is the case the figure
shows two possible simple-minded analytic models of the expected ratio
between the photometric discovered planets and the ones detected by RV
observations, one with $P^{-1/3}$ and the other with $P^{-2/3}$
dependence. The latter presents the geometrical effect of the
transiting planets only, while the former tries to take into account
the selection effect of the RV technique too, which is also less
effective for longer periods. To calculate the expected number of RV
detections we need a much more detailed study, which takes into
account the mass-period distribution of the population of the planets
as a whole (Gaudi et al. 2005), beyond this short review. If we
naively assume that the RV detection efficiency is proportional to the
RV amplitude, which goes like $P^{-1/3}$, then we may deduce that the
expected ratio of detection should vary like $P^{-1/3}$ too. We
therefore expect the actual ratio of the two sets of planets to be
somewhere between the two dashed lines in the figure. The figure
suggests that the actual ratio is much smaller than expected for
periods of 3--6 days. This is consistent with the suggested selection
effect discussed in the previous section. We expect this ratio to
substantially improve with more photometric observations per star.
 
As the photometric searches become more and more efficient, more
RV resources are needed to identify the true transiting planets. As
these resources (e.g., Bouchy et al. 2006) are rather limited, the
bottleneck of the ground-based photometric searches might soon be the
RV follow-up observations (e.g., Pont et al. 2008).  The problem is
enhanced by the fact that the small-telescope projects choose their
exposure time so that their detection will be most effective for stars
of about 12--13 mag.  This
causes the transiting planets discovered by photometric search to be
relatively faint, and therefore more difficult to follow by RV
observations.

One way to overcome this problem is to set a follow-up program
consists of a few stages. In the first stage the candidates would be
observed by a relatively low-resolution spectrograph (e.g., Latham
1992) to weed out obvious false candidates. Most of the short-period
binaries and the blends would be detected and rejected in this early
stage of the follow-up program.  Only the good candidates would then
be observed with high-resolution spectrographs to prove the planetary
nature of the unseen companions and to derive their masses. Such a
mode of operation is already being performed in a few ground-based
systematic searches (e.g., Bakos et al. 2007).

Finally, this review concentrated on the ground-based searches
only. However, two large systematic space-borne searches are on their
way. CoRoT is already working for more than a year (see Baglin et
al. in this volume) and producing superb lightcurves of newly
discovered planets (e.g., Alonso et al., 2008; Aigrain et al. 2008;
Moutou et al. 2008), and Kepler is about to be launched in less than a
year (Borucki et al 2008). We expect that with their superb precision
these two satellites will find more transiting planets in the range of
Neptune and even super-Earth size.

\section*{Acknowledgements}
I heartily thank C. Alonso, G. Bakos, A. Cameron, P. Kabath,
P. McCullough, F. O'Donovan, P. Sackett and A. Udalski who were kind
enough to answer my questions about the technical details of their
projects. I thank F. Pont for being patient with the manuscript of the
paper and O. Tamuz for critical reading of the manuscript. This work
was supported by the Israeli Science Foundation. I am indebted to the
Radcliffe institute for advanced studies at Harvard, where the last
version of the paper was prepared.




\begin{thebibliography}{}

\bibitem[]{} {Afonso C. et al} 2006, 
\textit{IAU Colloquium no. 200,
Direct Imaging of Exoplanets: Science \& Techniques}, 79

\bibitem[]{} {Afonso, C. \& Henning, Th.} 2007, \textit{Transiting
Extrapolar Planets Workshop, ASP Conference Series, Vol. 366}, 326

\bibitem[]{}
{Aigrain, S. et al.} 2008
\textit{A\&A}, 488, 43

\bibitem[]{}
{Alonso, R. et al.} 2004
\textit{ApJ}, 613, 153

\bibitem[]{}
{Alonso, R. et al.} 2008
\textit{A\&A}, 482, 21

\bibitem[Bakos et al. (2002)]{Bakos02}
{Bakos, G., Kov\'acs, G., \& Noyes, R.} 2005,
\textit{MNRAS}, 356, 557

\bibitem[Bakos et al. (2004)]{Bakos04}
{Bakos et al.} 2004,
\textit{PASP}, 116, 266

\bibitem[Bakos et al. (2007)]{Bakos07}
{Bakos et al.} 2007,
\textit{ApJ}, 670, 826

\bibitem[]{}
{Barge, P. et al.} 2008
\textit{A\&A}, 482, 17

\bibitem[]{} 
{Bayliss D.D.R., \& Sackett, P.D.} 2007, 
\textit{Transiting Extrapolar
Planets Workshop, ASP Conference Series, Vol. 366}, 320

\bibitem[]{} 
{Borucki et al.} 2007, 
\textit{Exoplanets: Detection, Formation and Dynamics, Proceedings of
  the International Astronomical Union, IAU Symposium, Volume 249}, 17

\bibitem[Bouchy \& the SOPHIE team (2006)]{Bouchy06}
{Bouchy \& the SOPHIE team} 2006,
\textit{Tenth Anniversary of 51 Peg-b: Status of and prospects for hot Jupiter studies}, 319

\bibitem[]{char00}
{Butler et al.} 2004,
\textit{ApJ}, 617, 580

\bibitem[]{}
{Charbonneau et al.} 2000,
\textit{ApJ}, 529, L45

\bibitem[]{}
{Charbonneau et al.} 2002,
\textit{ApJ}, 568, 377

\bibitem[]{}
{Gaudi, B., Seager, S., \& Mallen-Ornelas, G.} 2005,
\textit{ApJ}, 623, 472

\bibitem[]{gill07}
{Gillon, M. et al.} 2007,
\textit{A\&A}, 472, 13

\bibitem[Henry et al. 2000]{hen00}
{Henry, G. W., Marcy, G. W., Butler, R. P., \& Vogt, S. S.} 2000,
\textit{ApJ}, 529, L41

\bibitem[]{} {Horne, K.} 2007, \textit{Scientific Frontiers in
Research on Extrasolar Planets, ASP Conference Series, Vol 294}, 361

\bibitem[]{} {Kabath, P.} 2007, \textit{Transiting
Extrapolar Planets Workshop, ASP Conference Series, Vol. 366}, 23

\bibitem[]{}
{Kov\'acs, G., Zucker, S., \& Mazeh, T.} 2002,
\textit{A\&A}, 391, 396

\bibitem[Latham (1992)]{Latham92} {Latham} 1992, \textit{IAU
Colloq.~135, Complementary Approaches to Double and Multiple Star
Research}, 110

\bibitem[]{char00}
{Maness et al.} 2007,
\textit{PASP}, 119, 90

\bibitem[]{mazeh00}
{Mazeh, T. et al.} 2000,
\textit{ApJ}, 532, L55

\bibitem[]{mazeh05}
{Mazeh, T., Zucker, S., \& Pont, F.} 2005,
\textit{MNRAS}, 356, 955

\bibitem[]{}
{McCullogh, P.R. et al.} 2005
\textit{PASP}, 117, 783


\bibitem[]{}
{Moutou, C. et al.} 2008
\textit{A\&A}, 488, 47

\bibitem[]{}
{Pollacco, D. et al.} 2006
\textit{Ap\&AA}, 304, 253

\bibitem[]{}
{Pont, F. et al.} 2006
\textit{MNRAS}, 373, 231

\bibitem[]{}
{Pont, F. et al.} 2007a
\textit{A\&A}, 487, 749

\bibitem[]{} {Pont F. et al} 2007b, 
\textit{Transiting Extrapolar
Planets Workshop, ASP Conference Series, Vol. 366}, 3

\bibitem[]{}
{Pont, F. et al.} 2008
\textit{A\&A}, 465, 1069

\bibitem[]{}
{Queloz, D. et al.} 2000
\textit{A\&A}, 359, 13

\bibitem[Sato el al. 2005]{sato05}
{Sato et al. } 2005,
\textit{ApJ}, 633, 465

\bibitem[Tamuz et al. 2005]{sato05}
{Tamuz, O., Mazeh, T., \& Zucker, S.} 2005,
\textit{MNRAS}, 356, 1466


\bibitem[Udalski]{sato05}
{Udalski et al. } 2002a,
\textit{AcA}, 52, 1

\bibitem[Sato el al. 2005]{sato05}
{Udalski et al. } 2002b,
\textit{AcA}, 52, 115

\bibitem[Sato el al. 2005]{sato05}
{Udalski et al. } 2003,
\textit{AcA}, 53, 133

\bibitem[Udalski]{sato05}
{Udalski et al. } 2004,
\textit{AcA}, 54, 313

\bibitem[]{}
{Udalski et al. } 2005,
\textit{AcA}, 52, 115

\bibitem[]{}
{Vidal-Majar et al. } 2003,
\textit{Nature}, 422, 143

\bibitem[]{}
{Vidal-Majar et al. } 2004,
\textit{ApJ}, 604, L69


\bibitem[]{} {Weldrake, D.T.F.} 2007, \textit{Transiting Extrapolar
Planets Workshop, ASP Conference Series, Vol. 366}, 289


\end{thebibliography}
\end{document}